# On Novel Mechanism of a Pump Electromagnetic Wave Absolute Two-Plasmon Parametric Decay Instability Excitation in Tokamak ECRH Experiments

E.Z. Gusakov and A.Yu. Popov

*Ioffe Institute, St.-Petersburg, Russia*

Novel mechanism leading to excitation of absolute two – upper hybrid – plasmon parametric decay instability (TPDI) of a pump extraordinary (X) wave is discussed. It is shown that the upper hybrid (UH) plasmon can be 3D trapped in the presence of both a non-monotonous density profile and a finite-size pump beam in a plane perpendicular to the plasma inhomogeneity direction. This leads to excitation of the absolute TPDI of the pump X wave, which manifests itself in temporal exponential growth of the trapped daughter UH wave amplitude and is perhaps the most dangerous instability for mm-waves, widely utilized nowadays in tokamak and stellarators for local plasma heating and current drive and being considered for application in ITER experiments.

*Introduction.* – The two-plasmon decay instability (TPDI) [1 - 3] in laser-produced plasmas is of interest both as a potential source of hot electrons [4] and as a source of half-harmonic light which may serve as a coronal temperature diagnostic [5, 6]. In fusion plasmas the TPDIs even in the presence of the high power microwaves (up to 1 MW in a single beam) being usually utilized in electron cyclotron resonance heating (ECRH) experiments are believed for a long while to be deeply suppressed by plasma inhomogeneity due to the convective loss effect firstly predicted in the pioneering papers by Profs. A.D. Piliya and M.N. Rosenbluth [7 - 9]. The further progress of the inhomogeneous plasma PDI theory, a corner - stone of which is a monotonic spatial variation of plasma parameters, were summarized in review [10] and presented in text-book [11]. The typical microwave power threshold of the TPDIs in magnetized plasmas, which can be derived within the model of a monotonous background plasma density profile, is in the range of 10 MW [12 - 14]. However, in the presence of a non-monotonous plasma density profile, often observed in the ECRH experiments and originated due to different physical mechanisms among which are features of plasma confinement in the magnetic island [15] or the electron pump-out effect manifesting itself as a consequence of an anomalous convective particle flux from the ECR layer at the intensive ECRH [16] the trapping of one or both daughter waves in the direction of a plasma inhomogeneity is recently shown analytically to be possible [17 - 20]. This trapping leads to full suppression of the daughter wave energy loss from the decay region in this direction and can be responsible for excitation of several convective and absolute TPDIs, the thresholds of which are drastically less than the ones predicted by the standard theory [12 - 14]. The excitation of the most dangerous absolute TPDIs can explain the anomalous phenomena observed last decade experimentally in the ECRH experiments: first, fast ion generation at the TJ-II stellarator [21] and the TCV tokamak [22] during the ECRH pulse; second, observations of anomalous backscattering of the pump wave in the presence of the magnetic island at the Textor [23, 24] and ASDEX-UG [25, 26] tokamaks. These phenomena can be interpreted as a consequence of the pump electromagnetic wave absolute TPDI excitation in case of the daughter wave/ waves trapping in an artificial 3D plasma wave-guide. The mechanism of the daughter wave 3D trapping considered up to now is not universal, requiring the presence of either the drift wave eddies, filaments or blobs possessing the local density maximum and aligned with the magnetic field [18], or a specific value of the plasma density in the magnetic island slightly exceeding the UHR value for half a pump frequency [20]. In this paper, we address the novel mechanism being responsible for the 3D trapping of the daughter plasmon in the presence of both a nonmonotonous density profile and a finite spot of the pump beam in a plane perpendicular to the plasma inhomogeneity direction. The 3D trapping of the daughter upper hybrid can lead to excitation of the absolute TPDI (i.e. $t \to l_{UH} + l'_{UH}$) of the pump waves widely utilized nowadays in tokamak and stellarators for local plasma heating and current drive and being considered for application in ITER.

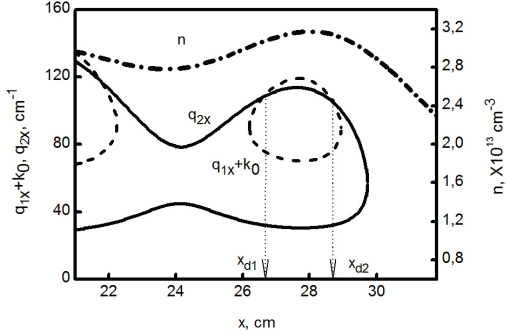

FIG. 1. (a, left and bottom axes) – The dispersion curves of the first UH wave up-shifted by the wave number of the pump X wave $q_{1x} + k_0$ (dashed curve) and the second UH wave (solid curve) are shown. In the points $x_{d1,2}$ the decay condition $k_0(x_{d1}) + q_{1x}(x_{d1}) - q_{2x}(x_{d1}) = 0$ is fulfilled; (b, right and bottom axes) – density profile (dash dotted curve) with the local maximum $x_m$ corresponding to the O-point of the magnetic island (2,1) [13].

*Basic equations.* − In this paper we consider the most simple but nevertheless relevant to an experiment the three-wave resonance interaction model and analyse the behaviour of the pump X wave propagating almost perpendicular to the magnetic field $\mathbf{H} = H\mathbf{e}_z$ in the density inhomogeneity direction along a unit vector $\mathbf{e}_x$ inwards plasma in the tokamak mid-plane:

$$\mathbf{E} = \mathbf{e}(\omega_0, x)\left[\frac{E_0(y,z)}{2}\sqrt{\frac{\omega_0}{ck_0(x)}}\exp\left(i\int^x k_0(x')dx' - i\omega_0 t\right) + c.c.\right], \quad (1)$$

where c.c. means complex conjugation of the first term, $\mathbf{e} = -i g/\varepsilon \mathbf{e}_x + \mathbf{e}_y$ is the polarization vector, $E_0(y,z) = \sqrt{8\pi/c}\sqrt{P_0/\pi w^2} \exp\left(-(y^2 + z^2)/2w^2\right)$ is the field amplitude, $P_0$ is the pump wave power, $w$ is the beam waist assumed constant as the domain of interest where the nonlinear interaction takes place is very small compared to the plasma size, and

$k_0 = \omega_0/c\sqrt{[\varepsilon^2 - g^2]/\varepsilon}$ is the radial wave-number with the components of the cold dielectric tensor $\varepsilon$ and $g$ being given explicitly below. The basic set of equations describing the X-mode pumping wave (1) decay into a couple of the electrostatic UH plasmons, propagating in opposite directions and represented by the potentials $\varphi_{1,2}(\mathbf{r},t) = \phi_{1,2}(\mathbf{r})\exp(\mp i\omega_{1,2}t)$, $\omega_2 = \omega_0 - \omega_1$, reads [20]

$$\begin{cases} \left[l_{T1}^2 \hat{q}_\perp^4 + \varepsilon(\omega_1)\hat{q}_\perp^2 - \omega_1^2/c^2\left(\varepsilon(\omega_1)^2 - g(\omega_1)^2\right) + \eta(\omega_1)\hat{q}_z^2\right]\phi_1(\mathbf{r}) = \sigma(\mathbf{r})\hat{q}_x^2 \phi_2(\mathbf{r}) \\ \left[l_{T2}^2 \hat{q}_\perp^4 + \varepsilon(\omega_2)\hat{q}_\perp^2 - \omega_2^2/c^2\left(\varepsilon(\omega_2)^2 - g(\omega_2)^2\right) + \eta(\omega_2)\hat{q}_z^2\right]\phi_2(\mathbf{r}) = \sigma^*(\mathbf{r})\hat{q}_x^2 \phi_1(\mathbf{r}) \end{cases} \quad (2)$$

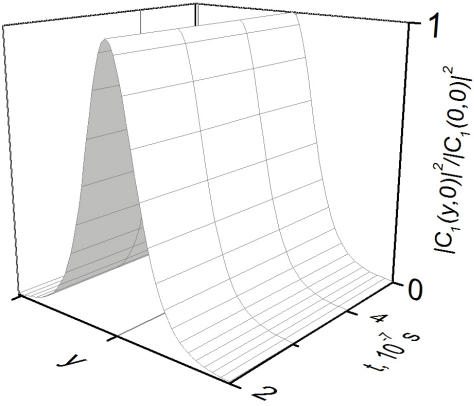

FIG. 2. Result of numerical evaluation of the partial differential equation (4) - $|C_1(t,y,0)|^2/|C_1(t,0,0)|^2$, $P_0 = 400kW$.

where $l_{T1,2}^2 = \dfrac{3}{2}\dfrac{v_{te}^2}{4\omega_{ce}^2 - \omega_{1,2}^2}\dfrac{\omega_{pe}^2}{\omega_{1,2}^2 - \omega_{ce}^2}$, $\hat{q}_\perp^2 = -\dfrac{\partial^2}{\partial x^2} - \dfrac{\partial^2}{\partial y^2}$,

$\hat{q}_z^2 = -\dfrac{\partial^2}{\partial z^2}$, $\sigma = \dfrac{3E_0}{2H}\dfrac{\omega_{pe}^2|\omega_{ce}|^2\omega_0^2}{(\omega_0^2 - \omega_{ce}^2)(\omega_1^2 - \omega_{ce}^2)(\omega_2^2 - \omega_{ce}^2)}$,

the components of the cold dielectric tensor are

$\varepsilon = 1 - \dfrac{\omega_{pe}^2}{\omega_{1,2}^2 - \omega_{ce}^2}$, $g = \dfrac{|\omega_{ce}|}{\omega_{1,2}}\dfrac{\omega_{pe}^2}{\omega_{1,2}^2 - \omega_{ce}^2}$, $\eta = 1 - \dfrac{\omega_{pe}^2}{\omega_{1,2}^2}$ and

$\omega_2 = \omega_0 - \omega_1$. Following an approach proposed in [27], in the first step we neglect the non-linear pumping (i.e. the RHS of equations (2)) and seek a solution of the homogeneous equations by means of the WKB approach which yields $\phi_{1,2}(x) = C_{1,2}f_{1,2}(x)$, $C_{1,2} = const$,

$$f_s(x) = \left|D_{sq}\left(q_{sx}^+(x)\right)\right|^{-1/2}\exp\left(i\int^x q_{sx}^+(\xi)d\xi\right) + \left|D_{sq}\left(q_{sx}^-(x)\right)\right|^{-1/2}\exp\left(i\int^x q_{sx}^-(\xi)d\xi\right), s = 1,2, (3)$$

and $q_{sx}^\pm = \sqrt{-\varepsilon \mp \sqrt{\varepsilon^2 + \omega_s^2/c^2(\varepsilon^2 - g^2)l_{Ts}^2}}/(\sqrt{2}l_{Ts})$. Far from the UHR $q_{sx}^+$ and $q_{sx}^-$ correspond to the «warm» and «cold» branches of the UH wave's dispersion curve at $q_{y,z} = 0$. The factors $\propto |D_{sq}|^{-1/2}$ in (3) where $D_{sq}(q_{sx}) = 2q_{sx}(2l_{Ts}^2 q_{sx}^2 + \varepsilon)$ provide the conservation of both the UH waves power fluxes in the direction of the plasma inhomogeneity. The UH wave can be trapped in the radial direction in a vicinity of the local maximum of the density profile $x_m$ corresponding to the O-point of the magnetic island [15] when it frequency obeys the quantization condition $\int_{x_{1l}^*}^{x_{1r}^*} q_{1x}^+(\omega_1^m,\xi)d\xi + \int_{x_{1r}^*}^{x_{1l}^*} q_{1x}^-(\omega_1^m,\xi)d\xi = \pi(2m+1)$ [20]. The coordinates $(x_{1l}^*, x_{1r}^*)$ corresponds to the UH wave turning points in the radial direction. In figure 1 for typical conditions of the TEXTOR tokamak ($T_e = 500eV$, $H_0 = 21000Gs$, $\omega_0/2\pi = 140GHz$) the dispersion curves $q_{1x}^\pm = q_{1x}^\pm(\omega_1^m, x)$ and $q_{2x}^\pm = q_{2x}^\pm(\omega_0 - \omega_1^m, x)$ at $q_{y,z} = 0$, as well as the actual radial density profile [15], are depicted. The dashed curve corresponds to the first daughter UH wave ($\omega_1/2\pi = 70.47GHz$) radial vector up-shifted by the radial wave-vector of the X wave $q_{1x} + k_0$; the solid curve corresponds to the dispersion curve of the second daughter UH wave ($\omega_2/2\pi = 69.53GHz$). In the point $x_d$ where the dashed and solid curves intersect the decay resonance condition $k_0(x_d) + q_{1x}^+(x_d) - q_{2x}^+(x_d) = 0$ is fulfilled. At the next step of the perturbation theory procedure we take into account the UH plasmons nonlinear pumping. This leads to the amplitudes $C_{1,2}$ of the potentials $\phi_{1,2}(\mathbf{r})$ being no longer constant. We neglect the diffraction energy loss of the second UH plasmon from the pump beam spot compared to its energy convection from the decay layer in the radial direction. Integrating the second equation in (2) gives the amplitude $C_2$ in a form

$$C_2(\mathbf{r}) = iC_1(y,z)\int_{\infty}^{x}\frac{ds\,\sigma^*(\zeta,y,z)}{\left|D_{2q}(\zeta)\right|^{1/2}}\exp\left(-i\int_{\infty}^{\zeta}q_{2x}^+(\xi)d\xi\right)\hat{q}_x^2 f_1(\zeta).$$

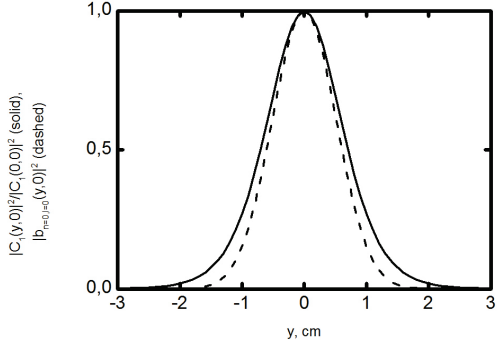

FIG. 3. $\left|C_1(y,0)\right|^2\big/\left|C_1(0,0)\right|^2$ (solid line), $\left|b_{n=0,l=0}(y,0)\right|^2$ (dashed line); $P_0 = 400kW$.

Substituting it into the first equation of (2), multiplying the latter by $f_1^*(m,x)$ and then integrating over $x$ we finally get

$$\left[\frac{\partial}{\partial t} + i\Lambda_y\frac{\partial^2}{\partial y^2} + i\Lambda_z\frac{\partial^2}{\partial z^2}\right]C_1(t,y,z) = \nu(y,z)C_1(t,y,z) \tag{4}$$

where $\Lambda_y = <l_{T1}^2 q_{1x}^2 + \varepsilon/2>/<q_{1x}^2\omega_1/\omega_{pe}^2>$, $\Lambda_z = <\eta>/<q_{1x}^2\omega_1/\omega_{pe}^2>$ are the coefficients averaged over the first UH plasmon radial localization area and describing the diffraction energy loss from the decay region,

$$\nu(y,z) = \frac{\left|\sigma(x_m,y,z)\right|^2}{<q_{1x}^2\omega_1/\omega_{pe}^2>}\frac{(l_{r1}^2 + l_{r2}^2)\left[q_{1x}^+(x_m)q_{2x}^+(x_m)\right]^2}{\int_{x_{1l}^*}^{x_{1r}^*}dx\left(\frac{\left|D_{1q}(q_{1x}^+(x_m))\right|\left|D_{2q}(q_{2x}^+(x_m))\right|}{\left|D_{1q}(q_{1x}^+(x))\right|} + \frac{\left|D_{1q}(q_{1x}^+(x_m))\right|\left|D_{2q}(q_{2x}^+(x_m))\right|}{\left|D_{1q}(q_{1x}^-(x))\right|}\right)}$$

is a term describing the non-linear pumping, $x_m$ is a coordinate of the O-point, $l_{r1,2} = \sqrt{2\pi}\left|\partial(q_{2x}^+ + k_0 - q_{1x}^+)/\partial x\right|_{x_{d1,2}}^{-1/2}$ is a length of the layer, within which the parametric decay

occurs, $x_d$ is a point, at which the decay resonance condition $k_0(x_{d1,2}) + q_{1x}^+(x_{d1,2}) - q_{2x}^+(x_{d1,2}) = 0$ is fulfilled (see fig.1), and the averaging procedure is defined as follows:

$$<f(x,q_{1x})> = \int_{x_{1l}^*}^{x_{1r}^*} dx \left( \frac{f(x,q_{1x})|_{q_{1x}^+(x)}}{|D_{1q}(q_{1x}^+(x))|} + \frac{f(x,q_{1x})|_{q_{1x}^-(x)}}{|D_{1q}(q_{1x}^-(x))|} \right) \bigg/ \int_{x_{1l}^*}^{x_{1r}^*} dx \left( \frac{1}{|D_{1q}(q_{1x}^+(x))|} + \frac{1}{|D_{jq}(q_{1x}^-(x))|} \right).$$

*Novel mechanism of the UH plasmon localization leading to excitation of absolute TPDI.* – The equation (4) describes evolution of the radially trapped UH plasmon in the presence of the non-linear pumping (RHS) and the energy loss (two last terms in LHS) from the decay region, manifesting itself in a form of diffraction. We will show that in the presence of the finite size pump beam (1) the UH plasmon can be localized within a spot of this beam. In its turn, this leads to an exponential growth of the UH plasmon amplitude. Then, we seek a temporally growing solution of (4): $C_1 = b(y,z)\exp(\gamma t)$. If $\gamma$ is large enough, i.e. $\gamma \leq |v(0,0)|$, the RHS of (4) can be expanded $v(y,z) - \gamma \approx v(0,0) - \gamma - v(0,0)(y^2 + z^2)/w^2$ that reduces (4) to

$$\left[ \Lambda_y \frac{\partial^2}{\partial y^2} + \Lambda_z \frac{\partial^2}{\partial z^2} - i(v(0,0) - \gamma) + iv(0,0)\frac{y^2 + z^2}{w^2} \right] b = 0 \tag{5}$$

The solution of (5) being given in terms of the Hermitian polynomials

$$b_{n,l}(y,z) = \exp\left\{ -\exp\left(i\frac{\pi}{4}\right) \sqrt[2]{\frac{|v(0,0)|}{\Lambda_y}} \frac{y^2}{w} \right\} H_n \left\{ \exp\left(i\frac{\pi}{8}\right) \sqrt[4]{\frac{|v(0,0)|}{\Lambda_y}} \frac{y}{\sqrt{w}} \right\}$$

$$\times \exp\left\{ -\exp\left(i\frac{\pi}{4}\right) \sqrt[2]{\frac{|v(0,0)|}{\Lambda_z}} \frac{y^2}{w} \right\} H_l \left\{ \exp\left(i\frac{\pi}{8}\right) \sqrt[4]{\frac{|v(0,0)|}{\Lambda_z}} \frac{y}{\sqrt{w}} \right\} \tag{6}$$

demonstrates the possibility of the UH plasmon capture within a spot of the pump beam. Thus, the exponentially growing solution of (4) with the growth rate being determined as follows

$$\gamma_{n,l} = v(0,0) - \left[ (2n+1)\sqrt{\frac{|v(0,0)|\Lambda_y}{2w^2}} + (2l+1)\sqrt{\frac{|v(0,0)|\Lambda_z}{2w^2}} \right] \tag{7}$$

is solely possible when the eigen-modes (6) due to the finite-size pump beam (1) are excited. The eigen-frequency corresponding to (6) and being a correction term to the eigen frequency $\omega_1^m$ of the radial eigenmode $f_1(m,x)$ is

$$\delta\omega_1^{n,l} = (2n+1)\sqrt{\frac{|v(0,0)|\Lambda_y}{2w^2}} + (2l+1)\sqrt{\frac{|v(0,0)|\Lambda_z}{2w^2}} \tag{8}$$

For the most dangerous fundamental mode $\delta\omega_1^{0,0}$ is much less than the frequency shift between the nearby eigen frequencies $\left|\omega_1^m - \omega_1^{m-1}\right|$ thus justifying the procedure of the equation (4) derivation. The threshold of the absolute TPDI excitation is determined by the equation

$$\nu\left(P_{th}^{n,l};0,0\right) = \frac{\left[(2n+1)\sqrt{\Lambda_y} + (2l+1)\sqrt{\Lambda_z}\right]^2}{2w^2} \quad (9)$$

In fig.2 the result of numerical evaluation of equation (4) for the pump power $P_0 = 400kW$ and the typical TEXTOR experimental parameters is presented. It demonstrates the excitation of the eigenmode-like structure, the $y$ - distribution $|C_1(t,y,0)|^2/|C_1(t,0,0)|^2$ of which retains its shape conserved in time. The $z$ - distribution of the solution $|C_1(t,0,z)|^2/|C_1(t,0,0)|^2$ keeps its shape conserved as well. In fig.3 the distribution $|C_1(y,0)|^2/|C_1(0,0)|^2$ at a fixed moment of time along with the fundamental mode squared $|b_{n=0,l=0}(y,0)|^2$ (dashed line) being a solution of the reduced differential equation (5) is illustrated. A reasonable agreement between these distributions is shown. In fig.4 the growth rate of the 3D trapped UH plasmon derived by numerical evaluation of the equation (4) (scatter symbols) and the analytical dependence (7) given by the solid line are shown. As we can see the dependences do match perfectly. In this case the power threshold of the absolute TPDI excitation is $P_{th}^{0,0} = 4kW$ whereas the growth rate at the power level of 400 - 600 kW usually utilized in Textor tokamak ECRH experiments is extremely high ($2\gamma \simeq 6 \div 8 \times 10^7 s^{-1}$) to complicate quasi linear instability saturation by variation of plasma parameters due to MHD or transport phenomena.

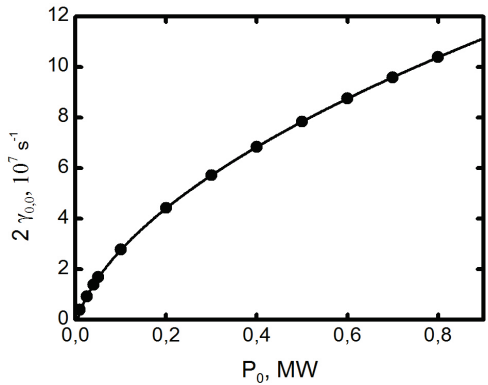

FIG. 4. The growth rate $2\gamma_{0,0}$. Scatter symbols – the result of numerical evaluation of equation (4). Solid line – equation (7); $P_{th}^{0,0} = 4kW$.

*Conclusions.* – In the paper we introduce the novel mechanism of the pump X wave absolute TPDI excitation. We have shown that if the trapping of the daughter plasmon in the direction of plasma inhomogeneity occurs, no matter which nature this phenomenon is, it can be also localized on the plane perpendicular to the plasma inhomogeneity direction due to the finite spot of the pump beam. The 3D trapping of the daughter upper hybrid plasmon leads to excitation of the low-threshold absolute TPDI of the pump X wave, the growth rate of which is drastically high $2\gamma \simeq 6 \div 8 \times 10^7 s^{-1}$. The mechanism discussed in the paper can be responsible for excitation of the absolute TPDI $t \to l_{EB} + l_{IB}$ (IB, EB – ion/electron Bernstein waves) in the TCV tokamak. The growth rate of this instability thoughts to be much higher than the one considered in [17] when the EB plasmon toroidal localization is provided by the effect of tokamak toroidal periodicity, and serves as a main candidate for explanation of fast ion acceleration observed in experiment [21, 22]. Thus, the different scenarios of the absolute TPDI ($t \to l_{UH} + l'_{UH}$ or $t \to l_{EB} + l_{IB}$) excited by the universal mechanism considered in the paper can explain the anomalous phenomena observed at ECRH experiments in different machines and seems to be the most dangerous for mm-waves, widely utilized nowadays in tokamak and stellarators for local plasma heating and current drive and being considered for application in ITER experiments.